\documentclass[conference]{IEEEtran}
\usepackage{hyperref}
\IEEEoverridecommandlockouts
\usepackage{subfig}
\usepackage{diagbox}
\usepackage{balance}
\usepackage{graphicx}
\usepackage{textcomp}
\usepackage{xcolor}
\usepackage{caption}
\usepackage{makecell}
\usepackage{booktabs}

\def\BibTeX{{\rm B\kern-.05em{\sc i\kern-.025em b}\kern-.08em
    T\kern-.1667em\lower.7ex\hbox{E}\kern-.125emX}}
\usepackage{fancyhdr}

\begin{document}

\title{MRSch:  Multi-Resource Scheduling for HPC}
\author{\fontsize{11}{11}\selectfont %
{
Boyang Li\IEEEauthorrefmark{1},
Yuping Fan\IEEEauthorrefmark{1}, 
Matthew Dearing\IEEEauthorrefmark{1},
Zhiling Lan\IEEEauthorrefmark{1}\thanks{Zhiling Lan's current affiliation is University of Illinois Chicago, and her current contact is \ttfamily\upshape{zlan@uic.edu}.}, 
Paul Rich\IEEEauthorrefmark{2}, 
William Allcock\IEEEauthorrefmark{2}, 
Michael Papka\IEEEauthorrefmark{2},\IEEEauthorrefmark{3}
}
\vspace{1mm}\\
\fontsize{10}{10}\selectfont\itshape
\IEEEauthorrefmark{1}Department of Computer Science, 
Illinois Institute of Technology, Chicago, IL, USA\\
\fontsize{9}{9}\selectfont\ttfamily\upshape
\{bli70, yfan22, mdearing\}@hawk.iit.edu, lan@iit.edu

\vspace{1mm}\\
\fontsize{10}{10}\selectfont\rmfamily\itshape
\IEEEauthorrefmark{2} Argonne National Laboratory, Lemont, IL , USA\\
\fontsize{9}{9}\selectfont\ttfamily\upshape
{richp,allcock,papka}@anl.gov

\vspace{1mm}\\
\fontsize{10}{10}\selectfont\rmfamily\itshape
\IEEEauthorrefmark{3} Northern Illinois University, IL,  USA\\
\fontsize{9}{9}\selectfont\ttfamily\upshape

}

\maketitle

\thispagestyle{fancy}
\lhead{}
\rhead{}
\chead{}
\lfoot{\footnotesize{
IEEE Cluster 2022, September 6-9, 2022, Heidelberg, Germany
\newline  978-1-6654-9856-2/22/\$31.00 \copyright 2022 IEEE}}
\rfoot{}
\cfoot{}
\renewcommand{\headrulewidth}{0pt}
\renewcommand{\footrulewidth}{0pt}

\begin{abstract}
Emerging workloads in high-performance computing (HPC) are embracing significant changes, such as having diverse resource requirements instead of being CPU-centric. This advancement forces cluster schedulers to consider multiple schedulable resources during decision-making. Existing scheduling studies rely on heuristic or  optimization methods, which are limited by an inability to adapt to new scenarios for ensuring long-term scheduling performance. We present an intelligent scheduling agent named MRSch for multi-resource scheduling in HPC that leverages direct future prediction (DFP), an advanced multi-objective reinforcement learning algorithm. While DFP demonstrated outstanding performance in a gaming competition, it has not been previously explored in the context of HPC scheduling.  Several key techniques are developed in this study to tackle the challenges involved in multi-resource scheduling.  These techniques enable MRSch to learn an appropriate scheduling policy automatically and dynamically adapt its policy in response to workload changes via dynamic resource prioritizing. We compare MRSch with existing scheduling methods through  extensive trace-base simulations. Our results demonstrate that MRSch improves scheduling performance by up to 48\% compared to the existing scheduling methods.    
\end{abstract}

%
%
%

%

%

\maketitle

\section{Introduction} \label{sec:intro}
The cluster scheduler, also known as a batch scheduler,  plays a critical role in high-performance computing (HPC), with the responsibility of determining the order in which jobs are executed. Existing cluster schedulers are CPU-centric. However, exponential growth in computing power has enabled HPC systems to tackle much more complex scientific problems. These emerging workloads have diverse resource requirements beyond the CPU.  For example, I/O intensive applications can take advantage of a burst buffer with dramatically improved performance \cite{kougkas2016leveraging}. For these applications, raw CPU power is not necessarily the primary resource that determines performance, but the allocation with respect to fast storage is more crucial. Such a change requires the scheduler to consider \textit{multi-resource scheduling} where the scheduling problem is to optimize the use of multiple schedulable resources, e.g., CPU, burst buffer, power, and so on.

\begin{figure*} 
    \centering
  \subfloat[Job waiting queue.]{%
   \centering
      \hspace{6mm} 
      \includegraphics[width=0.35\linewidth]{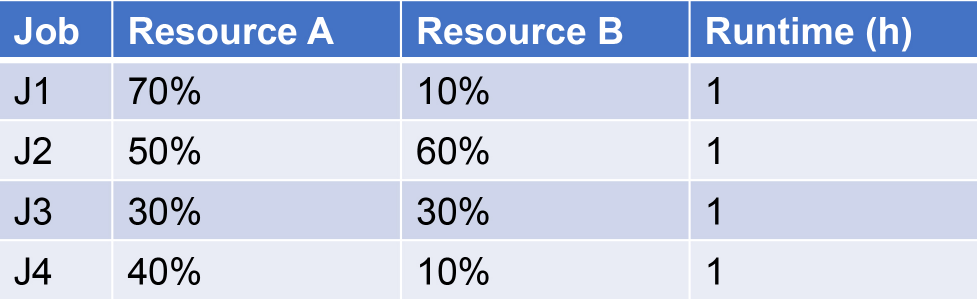}}
    \label{df}\hfill
  \subfloat[Scheduling decisions by different methods.]{%
        \includegraphics[width=0.4\linewidth]{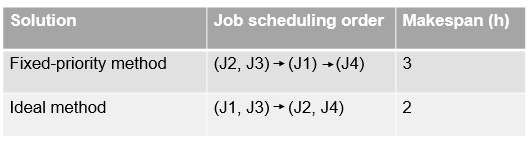}}
        \label{am}
\caption{An example illustrating the limitation when fixing  the  priority of  each  objective  for  job  scheduling.}
\label{fig:example}
 \vspace{-0.5cm}
\end{figure*}

Existing multi-resource scheduling methods often rely on heuristics  \cite{ghodsi2011dominant,grandl2014multi,delimitrou2014quasar,grandl2016altruistic}. Among them, dominant resource fairness (DRF) \cite{ghodsi2011dominant} and Tetris \cite{grandl2014multi} are widely cited. While these heuristics have been demonstrated to be effective for the workloads in data centers, they are not suitable for multi-resource scheduling in HPC because these  two communities adopt different computing modes and target very different workloads. For example, in DRF, each job consists of multiple tasks, and the scheduling process is to determine the proper number of tasks per job (i.e., a malleable job) to maximize the minimum dominant share among jobs. In contrast, HPC is dominated by rigid parallel jobs with a fixed number of tasks. A key feature of HPC scheduling is to improve resource utilization while preventing job starvation where large-sized or long-running jobs are perpetually held in a waiting queue.  

A few research studies presented heuristic or classical optimization methods for multi-resource scheduling in HPC. Sun et al. discussed list scheduling and pack scheduling, both being proposed for scheduling moldable jobs \cite{garey1975bounds,sun2018scheduling}. One variant of the list scheduling method extends first-come, first-serve (FCFS) to multi-resource scheduling \cite{mu2001utilization}. While heuristic methods are fast, they cannot deliver an optimal solution to a scheduling problem. Optimization methods were also explored for multi-resource scheduling \cite{wallace2016data,yang2013integrating,wierman2008stochastic,ren2012provably,fan2019scheduling}. These methods formulate the scheduling problem into a single-objective or multi-objective optimization problem. Studies suggested that the optimization-based methods, especially the multi-objective optimization approach, result in better scheduling performance \cite{fan2019scheduling}. 

Recent efforts explored reinforcement learning (RL) for cluster scheduling \cite{mao2016resource, zhang2020rlscheduler, fan2021deep}. 
Distinguishing from heuristic and optimization methods that concentrate on the \emph{immediate effect}, reinforcement learning processes a sequence of decisions where each decision can impact the next. Through training, an RL agent learns to make an informative decision to optimize the \emph{long-term} effect resulting from each scheduling decision as a sequence of actions (e.g., effects on the current and future resource utilization) \cite{sutton2018reinforcement}. Moreover, a common drawback of heuristics and optimization methods is the \emph{lack of adaptation}.  An intriguing feature of RL is its ability to \emph{adapt its actions} automatically to dynamic changes in workloads or system states. As such, RL offers a promising direction for improving cluster scheduling. 
Also, existing RL-driven scheduling techniques mainly concentrate on single-resource scheduling.

In this study, we suggest that \textit{multi-objective reinforcement learning (MORL)} is a natural approach for multi-resource scheduling.  A simple approach may  extend existing RL-driven scheduling algorithms to multi-resource scheduling by using a scalar reward with a fixed priority per resource (e.g., assigning a weight of 50\% per resource for two-resource scheduling). However, such an  extension has a serious drawback. 
The following example illustrates the limitation of the fixed priority method. Consider a two-resource scheduling scenario, denoted by Resources A and B, with an empty initial system state. Four jobs in a waiting queue have a different resource demand expressed by the percentage of the system resource capacity needed, as shown in Figure 1(a).  
When using a fixed weight method (e.g.,  equally maximize the utilization of Resources A and B), the job scheduling order for this fixed weight method is (J2, J3), (J1), and (J4). As a result, the makespan (i.e., the time spent to  complete all jobs) is three hours.  However, the ideal job scheduling order should be (J1, J3) followed by (J2, J4), which results in a shorter makespan of only two hours. This example demonstrates that statically weighting multiple resources fails to provide an efficient scheduling. 

Inspired by recent advancements in MORL, we investigate a new approach for multi-resource scheduling by developing an intelligent scheduling agent named \emph{MRSch}. Our design leverages an advanced MORL method called \emph{direct future prediction (DFP)} \cite{dosovitskiy2016learning}, which was proposed by Intel for gaming. 
Similar to classic RL methods, a trained DFP agent can make an intelligent decision by considering the long-term effect of each scheduling decision in a sequence of actions.  
Distinguishing from classical RL using a scalar reward, DFP  dynamically prioritizes each objective at runtime after being properly set up with a goal vector (\S \ref{sec:vector setting}). Such an adaption is essential for dynamic resource changes experienced in multi-resource scheduling scenarios.  

While the inherent advantages of DFP are appealing, it cannot be directly applied to our problem, and several technical challenges must be overcome. The design of MRSch contains several core components that address these challenges.  
First, the original DFP algorithm uses an image input to encode each frame of a video game. Previous scheduling studies for data centers also used a fixed-size 2D image for encoding job and system information (i.e., one dimension for resource availability and the other for time duration) \cite{mao2016resource}. Unfortunately, an image-based state representation is not appropriate for HPC scheduling. Unlike the tasks in data centers with a fixed time duration, HPC jobs may take seconds, days, or weeks to complete. As such, image-based encoding cannot effectively address this wide range of job duration. Instead, MRSch incorporates \emph{a vector-based encoding mechanism} for effectively representing user jobs and system resources in a scalable way (\S \ref{sec:state representation}). 
Second,  a convolutional neural network (CNN) \cite{lecun1989backpropagation} is used in DFP for information processing.  While CNNs are suitable for spatial data, user jobs and system states do not contain many spatial relationships. Therefore, we adopt the \emph{multilayer perceptron (MLP)} \cite{rumelhart1985learning} in the design of MRSch. 
Third, an essential input to drive DFP is \emph{construction of a goal vector} that dynamically captures the relative priority of different resources.  The design of MRSch uses a simple yet effective technique to automatically adjust the weight of each resource preference so as to pay more attention to the highly demanding resource by user jobs (\S \ref{sec:vector setting}). 
Fourth, given the unique characteristics of HPC workloads, advanced resource reservation and backfilling are common features required for HPC for preventing job starvation and improving resource utilization. MRSch incorporates these \emph{HPC domain-specific techniques} by deploying a window-based reservation (\S \ref{sec:starvation}).  Finally, an efficient training strategy is leveraged by MRSch for fast convergence (\S \ref{sec:training}). 

Implemented in TensorFlow~\cite{MRSch}, we  evaluate MRSch by extensive trace-based simulations with real-world job traces collected from the Theta machine at the Argonne Leadership Computing Facility (ALCF) \cite{ALCF}.  To extensively evaluate MRSch under various resource confliction and saturation environments, we generate a series of workloads from these real traces that encompass a range of workloads. We compare MRSch with heuristic, classical optimization, and an extension of a single-objective RL.   

We consider a setup where an HPC system has up to \(R\) schedulable resources. For simplicity, we initially restrict our focus to two schedulable resources: CPU and burst buffer. Following this setting, a case study is presented to show that MRSch can be readily extended to additional schedulable resources. Our experiments conclude that MRSch outperforms the existing methods by up to 48\% with respect to overall scheduling performance.


\begin{table*}
\begin{center}
 \caption{Comparison of MRSch with existing multi-resource cluster scheduling methods.}

  \label{tab:comparison of scheduling policy}
 \begin{tabular}{|l|c|c|c|c|}
\hline
\diagbox[width = 16em]{\thead{Features}}{\thead{Methods}}& \thead{Heuristics \\
\cite{mu2001utilization,sun2018scheduling,ghodsi2011dominant,grandl2014multi,garey1975bounds} }& \thead{Classical optimization \\ \cite{wallace2016data,yang2013integrating,wierman2008stochastic,ren2012provably,fan2019scheduling}} & \thead{ Existing RL-driven scheduling \\ \cite{fan2021deep,zhang2020rlscheduler,baheri2020mars,mao2016resource}}& \thead{MRSch} \\
\hline
\thead{Long-term scheduling effect}&$\times$ & $\times$&$\sqrt{}$ & $\sqrt{}$   \\
\hline
\thead{Automatic policy tuning}&$\times$ & $\times$&$\sqrt{}$ & $\sqrt{}$ \\

\hline
 \thead{Dynamic resource prioritizing}&$\times$ & $\times$&$\times$ & $\sqrt{}$   \\
 \hline
\thead{Training requirement}&$\times$ & $\times$&$\sqrt{}$ & $\sqrt{}$   \\
\hline
\end{tabular}
\end{center}
\vspace{-0.5cm}
\end{table*}


\section{Related Work and Background} \label{sec:background}

\subsection{Related Work} \label{secc:HPC sch}

On typical HPC clusters, cluster scheduling is responsible for allocating resources and determining the order in which jobs are executed. When submitting a job, a user is required to provide the resources required by the job and an estimate of job runtime. Submitted jobs are stored and sorted in a waiting queue based on the facility's prioritization policy. The scheduler then determines \textit{when} and \textit{where} to execute these queued jobs~\cite{allcock2017experience}. 

Unlike scheduling in data centers, HPC scheduling has several salient features. In particular, HPC is dominated by tightly-coupled parallel applications. Hence, \emph{advanced job reservation} and \emph{backfilling} are commonly used for preventing job starvation and improving resource utilization \cite{mu2001utilization,allcock2017experience}. Job reservation holds resources for the job at the head of the waiting queue to prevent starvation.  Backfilling enables subsequent jobs to move ahead to utilize free resources appropriate for that job. A widely used strategy is EASY backfilling, which allows short jobs to skip ahead in the queue only if they do not delay the current job waiting at the head of the queue \cite{mu2001utilization}. 

Considerable studies have been conducted to improve cluster scheduling by leveraging machine learning. For instance, one active topic is forecasting job characteristics or user behaviors to improve cluster scheduling, such as reported in \cite{CMU2019} with a summary of the challenges and limitations of applying machine learning for job characteristic prediction.  Distinguishing from this research, in recent years several pioneering studies explored reinforcement learning for HPC scheduling (i.e., sequential decision making). For example, RLScheduler deployed a new kernel-based neural network structure
and trajectory filtering mechanism to stabilize the learning process \cite{zhang2020rlscheduler}. MARS combined heuristics and a deep RL actor-critic algorithm to optimize HPC systems for legacy and complex workflows \cite{baheri2020mars}. DRAS leveraged a hierarchical neural network that incorporates HPC-specific scheduling features \cite{fan2021deep}. These studies targeted CPU-only scheduling. 


For multi-resource scheduling, heuristic methods are commonly used, such as co-scheduling CPUs and memory in data centers \cite{ghodsi2011dominant,grandl2014multi,verma2015large,grandl2016altruistic}. Among them, dominant resource fairness (DRF) and Tetris are well-known methods \cite{ghodsi2011dominant,grandl2014multi}. DRF adopts a max-min fairness algorithm for the dominant resources to ensure that no user is better off if the resources, such as CPU and memory, are equally partitioned among them \cite{ghodsi2011dominant}. Tetris presents a multi-dimensional bin packing method that improves the average job completion time by preferentially serving jobs that have less remaining work compared to other jobs \cite{grandl2014multi}. These studies targeted typical workloads seen in data centers with jobs composed of multiple tasks and scheduling decisions designed to determine how many tasks for each job should be selected. 

Unfortunately, these techniques are not suitable for multi-resource scheduling in HPC for two reasons. First, the scheduling objective in HPC is to optimally schedule jobs in the waiting queue (instead of tasks within the jobs, as in data centers). Second, large-sized, long-running rigid jobs are common in HPC, and preventing their starvation in the waiting queue is a crucial scheduling requirement. 

Existing multi-resource scheduling approaches in HPC can be broadly classified as either heuristics- or optimization-based methods. In list scheduling \cite{garey1975bounds,sun2018scheduling},  jobs are first organized in a priority list and assigned in sequence to the earliest available resources. An extension of FCFS to multi-resource scheduling is an instance of list scheduling. 
Classical optimization methods have also been considered 
for multi-resource scheduling \cite{wallace2016data,yang2013integrating,wierman2008stochastic,ren2012provably}. Yuping et al. \cite{fan2019scheduling} developed a multi-resource scheduling algorithm to explore a Pareto set for decision-making.  Heuristic and optimization methods are similar in that decisions are made for the best immediate effect, such as maximizing resource utilization at the decision-making moment. However, considering only immediate consequences may lead to suboptimal  performance in the long term. 

MRSch differs from these prior studies in multiple aspects, as summarized in 
Table \ref{tab:comparison of scheduling policy}.

\begin{figure*}
\centering
\includegraphics[height=2.4in,width=6in]{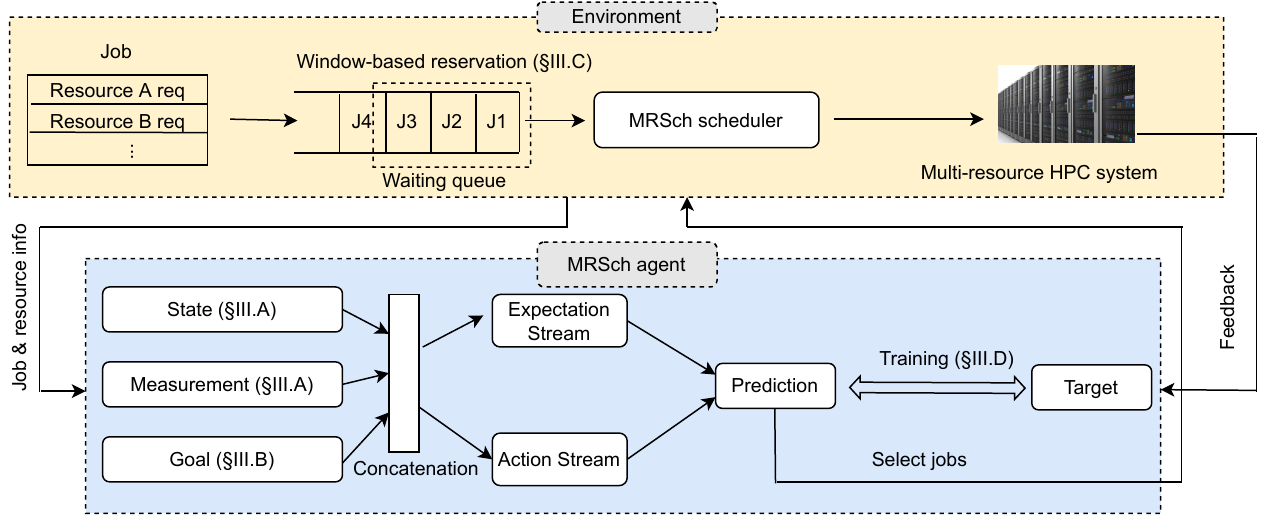}
\caption{Overview of MRSch. The environment (the top portion) denotes the HPC  multi-resource scheduling system. The MRSch agent (the bottom portion) contains three input modules (state, measurement, and goal) and interacts with the environment by observing environmental changes and making scheduling decisions (i.e.,  selecting jobs for execution). The arrows between the agent and the environment indicate the information flows between them.
}
\label{fig:dfp_design}
\end{figure*}

\subsection{Direct Future Prediction} \label{sec:DFP overreview}
Direct future prediction (DFP) is an advanced MORL algorithm developed in 2017~\cite{dosovitskiy2016learning}. Its foundational idea is to train an agent \emph{to predict the effect of different actions on future measurements, conditioned by the present state input, measurements, and goal.} DFP inherits the long-term scheduling impact of traditional reinforcement learning. Distinct from conventional RL with feedback as a scalar reward, feedback in DFP is in the form of a measurement (\emph{a vector}). Leveraging this extension, unlike traditional RL methods that learn a single objective according to a scalar reward, DFP can \emph{switch} goals (i.e., the product of the measurement and goal vector) under various circumstances. This switching is performed by dynamically adjusting the goal vector. 

DFP incorporates three input modules, each processing an image $s$ (i.e., a perception module), measurement $m$, and goal $g$ (i.e., reflecting the relative importance of each measurement) separately.  
The pursued objective can be expressed as a dot product of the predicted measurement change and goal vector.  The outputs of these modules are concatenated into a joint representation $j$ that is processed by two parallel streams, an expectation stream  and a normalized action stream, inspired by the dueling architecture introduced by DeepMind \cite{wang2016dueling}. These two streams are combined to produce a final prediction for each action. More details of DFP can be found in ~\cite{dosovitskiy2016learning}.

The DFP agent interacts with the environment to obtain the actual measurement change. The loss function between this measurement and the predicted measurement is used to train the neural network.  During training, the agent follows an $\epsilon$-greedy policy to avoid local optimums. During testing, the agent selects the action that yields the best-predicted outcome. 



\section{MRSch Design} \label{sec:MRSch Design}

 MRSch represents the scheduler as an intelligent agent that makes decisions for when and which jobs should be allocated to available resources  (Figure \ref{fig:dfp_design}).  \emph{The environment} includes job and resource information, along with system measurements, such as resource utilization. The objective of the MRSch agent is to maximize the utilization of each resource by taking the actions of selecting jobs for scheduling. Because resource scarcity dynamically changes, the weight per resource, represented by the \emph{goal} module, must adapt to dynamic environmental changes for optimizing job selection. 
 
The MRSch agent interacts with the environment over discrete scheduling instances. At a given instance, the agent reads the job and resource information as input for the \emph{state} and \emph{measurement} modules. The input of the goal module represents the weights of each measurement from the measurement module.  The outputs of these three modules are concatenated into a joint representation that is processed by the parallel \emph{expectation
stream} and  \emph{action stream}. The outputs of these streams are combined to produce a final \emph{prediction} of future measurements for each action.  The agent then takes an action by selecting jobs from the waiting queue  and obtains the actual future measurement (the \emph{target} module) fed back by the system.  MRSch trains the neural networks to improve the prediction accuracy of future measurements for each action by minimizing a loss function between the prediction and target. Key techniques designed into MRSch are described below.

\subsection{Input Modules} \label{sec:state representation}

The foremost challenge is formulating the specific HPC multi-resource scheduling problem as MORL. 
In the following, we describe our representations of the input modules featured in Figure \ref{fig:dfp_design}.

\textbf{State}. 
In the original DFP, the input of the state module is an image  \cite{dosovitskiy2016learning}. Encoding job and resource information as an image is not suitable for our case because it is difficult to capture critical  job  information (e.g., job  waiting time) in images. Instead, we encode the job and resource information as \emph{vectors}. 
Each waiting job is encoded as a vector of $(R+2)$ elements, where $R$ is the number of resources requested by the job, and the additional elements correspond to the user-supplied estimated runtime and queued time of the job.

For system resources, we encode each resource unit as a vector of two elements.
The first is a binary value representing resource availability (1 means available and 0 means not available). If the resource is occupied,  then we take the user-supplied runtime estimate and job start time to calculate this unit's estimated available time. The second element is the time difference between the resource unit's estimated available time and the current
time. If the resource is available, then we set this element to zero. The resource unit can be defined by the system administrator, e.g., a node for the CPU resource or a TB burst buffer as the unit for the burst buffer resource. Finally, we concatenate job information and resource information into a fixed-size vector as the input for the state module. 

Rather than using CNN within the state module as deployed in the original DFP,  we incorporate a multilayer perceptron (MLP). 
CNN works well on data with spatial relationships, such as image data \cite{alzubaidi2021review}. However, the features of our state input are independent. We show experimental results comparing MLP and CNN architectures in Section \S \ref{network performance}. 

We also use one neural network for all resources instead of one neural network per resource.  This design choice is based on two reasons. First, more training parameters are available for the state module with a single neural network configuration compared  to  separate  neural  networks.  Second,  if using multiple neural networks, job information would be encoded multiple times in the final joint representation, resulting in an inefficient redundancy. 

 Our state neural network consists of four layers, including the input layer, two fully-connected layers, and output layer. The input layer is connected to  two fully-connected layers activated by a leaky rectifier \cite{goodfellow2016deep}, and the second fully-connected layer is connected to the output layer.

\textbf{Measurement}. The inputs to the measurement module are the metrics of the scheduling objective. Different HPC facilities may have different scheduling objectives. A typical objective is to maximize the utilization for all schedulable resources. Suppose two types of resources, Resource A and B, are available,  and the site objective is to maximize the utilization of both resources. A measurement vector is defined as $<$\emph{Resource A util}, \emph{Resource B util}$>$, and a three-layer fully-connected network parses the measurement module. 

\textbf{Goal}. The values of the goal vector determine the weights of each measurement in the overall scheduling objective. 
Positive values correspond to maximizing the particular measurement, and negative values correspond to its minimization. Configuring the goal vector is described in the next subsection. 

\textbf{Action}. MRSch deploys a window to specify a range of jobs to select from within the waiting queue. 
Intuitively, the scheduler can select multiple jobs within this window
simultaneously, but this could result in an explosive number of actions.
Instead, MRSch decomposes a scheduling decision that includes several jobs in one action into a series of individual job selections.

\vspace{-0.1cm}
\subsection{Dynamic Resource Prioritizing} \label{sec:vector setting}
The fierceness of contention for each resource changes during multi-resource scheduling, so more consideration should be assigned to the more contentious resource. Therefore, dynamically adjusting the resource priority is essential.

In MRSch, dynamic resource priority is achieved by adjusting the \emph{goal vector} input to the goal module, $g$, that represents the weights of each measurement in the overall scheduling objective. A larger value of the goal vector means the corresponding measurement plays a more important role in the scheduling objective. MRSch gives preference to the resource with more fierce contentions.

\begin{table}
\centering
  \caption{Symbols and their descriptions.}
  \label{tab:Symbol}
  \begin{tabular}{cl}
    \toprule
   Symbol&Description\\
    \midrule
    N & number of jobs in the system.\\
    R & number of resources in the system. \\
    $t_{i}$ & user-supplied runtime estimate of job $i$ in waiting queue, \\
              
              ~&~remaining runtime estimate of job $i$ running on system.              \\
    $P_{ij}$ & percentage of requested resource $j$, \\
               ~&~(divided by the system resource $j$ capacity) for job $i$.\\
    $r_{j}$& goal vector value reflecting the contention fierceness \\
    ~&~of resource $j$ by all jobs, including running and queued.\\
 
        \bottomrule
\end{tabular}
\vspace{-0.5cm}
\end{table}

\begin{table*}
\centering
\caption{Workloads based on the production traces, representing light to heavy contention for the burst buffer.}

  \label{tab:workload}
\begin{tabular}{cccc}
    \toprule
  Workload &Number of requested nodes &Percentage of jobs requesting burst buffer & Burst buffer size range\\
    \midrule
S1& number of requested nodes in the trace & 50\% & [5 TB, 285 TB]\\
S2& number of requested nodes in the trace& 75\%&[5 TB, 285 TB]\\
S3& number of requested nodes in the trace&  50\%&[20 TB, 285 TB] \\ 
S4& number of requested nodes in the trace&  75\% & [20 TB, 285 TB] \\
S5& half of number of requested nodes in the trace&  75\% & [20 TB, 285 TB] \\
     \bottomrule
\end{tabular}
\vspace{-0.5cm}
\end{table*}

Suppose there are $R$ schedulable resources and the scheduling objective is to maximize resource utilization (Table \ref{tab:Symbol} lists all symbols and their corresponding meanings). MRSch sets the values in the goal vector as follows: 

\begin{equation}\label{normalizer}
r_{j}=\frac{\sum_{i=1}^{N} P_{ij} t_{i} }{\sum_{j=1}^{R}\sum_{i=1}^{N} P_{ij} t_{i}}
\end{equation}

Equation (\ref{normalizer}) describes how long (normalized) it will take to complete all the jobs' resource $j$ demands in the ideal situation where resource $j$ is fully utilized. A longer time represents a more fierce  resource contention. 

\subsection{Avoid Job Starvation} \label{sec:starvation}

HPC job sizes and runtimes can span broad scales in practice. A job size ranges from a single node to the entire HPC system comprised of thousands of compute nodes, and its runtime may vary from seconds to days. Such a variety in job characteristics presents a unique challenge for HPC scheduling: queued jobs, especially large-sized jobs, tend to be starved when small-sized jobs continue arriving into the queue and skip to the front while insufficient resources are available for the larger job. Directly applying  DFP to the multi-resource scheduling problem results in severe job starvation. 

MRSch adopts two techniques to overcome this challenge. First, a window-based design alleviates job starvation by providing higher priority to older jobs in the queue. Second, MRSch inherits the reservation strategy. At a given scheduling instance, the scheduler enforces a window at the front of the waiting queue.  When MRSch selects a job from this window, if its requested resources are available, then it is marked as ready and sent for immediate execution on the system. This
process repeats until the system no longer has sufficient available resources for the next job selected by the agent. This next job is then marked as reserved so that its requested resources will be held for its execution on the system at the earliest available time. In addition, EASY backfilling is leveraged to improve resource utilization. 

\subsection{Training Strategy} \label{sec:training}

To obtain a converged and accurate model for scheduling, the MRSch agent must gain experience through training from a large quantity of jobs with various job arrival patterns and diverse job characteristics. We train our MRSch agent with real workloads, along with sampled and synthetic workloads,  to increase its robustness toward workload changes. 

We follow the principle of gradual improvement to learn a robust model. MRSch begins with common represented cases and incrementally improves its capability with unseen rare cases. In particular, three types of job sets and a three-phase training process  are employed to train MRSch in the following order: a set of sampled jobs from real job traces, real job traces, and synthetic jobs generated to represent previously unseen  patterns. The sampled job sets have controlled job arrival rates that provide the easiest learning environment for MRSch to learn good scheduling decisions within a controlled environment. Subsequent training on real job traces with varying job arrival patterns enables MRSch to learn more complex scenarios. The final phase includes synthetic job sets to tune MRSch with experiences from a broader variety of potential states that may have not been seen during the first two sets. Results comparing different training strategies are presented in  \S \ref{training performance}.  

\section{Implementation and Evaluation} \label{sec:eval}

MRSch is implemented in TensorFlow~\cite{MRSch}. We evaluate MRSch through trace-based simulation using real workloads collected from a production system. In our experiments, MRSch interacts with CQSim, a trace-based HPC job scheduling simulator that has been used in various scheduling studies for a decade~\cite{CQSim}. A real system
processes jobs from user submissions, while CQSim imports
jobs by reading the job arrival information from a trace. The
simulator emulates system execution by advancing the simulation
clock according to the job runtime recorded in the trace.
Changes in the job wait queue or the system trigger the simulator
to send scheduling requests to the MRSch agent. Typical triggers include the submission of a new job to the queue or a 
running job leaving the system.

For simplicity of presentation, we first confine our attention to two resources and later present a case study featuring more resources in \S \ref{sec:3 resources}.

\subsection{Workload Traces} \label{sec:trace}

A variety of resources beyond CPUs may be considered as schedulable resources.  Given that the burst buffer is widely deployed in production supercomputers \cite{Trinity,Cori}, we evaluate MRSch with the scheduling of CPU and burst buffer. 

Our workload trace is a five-month historical job trace in 2018 from Theta at ALCF \cite{Theta}.  This trace only contains CPU request information, so we extend the data with burst buffer (BB) requests, assuming a shared burst buffer of 1.26 PB. To compensate for this lack of burst buffer information in the trace, a corresponding Darshan \cite{Darshan} trace extracts the amount of data moved between compute nodes and the parallel file system, which is then considered as the potential burst buffer request for each job. During the five months, 40\% of the jobs have Darshan I/O records, and 17.18\% have more than 1 GB of data transferred. The amount of transferred data is assigned as the corresponding job's burst buffer request, with a range of requested burst buffer sizes between 1 GB to 285 TB. A limitation is that the burst buffer was not heavily utilized during the time of this historical trace because the burst buffer was a relatively new resource, and not all applications had been refactored to benefit from this new feature. Also, some jobs did not include Darshan I/O recordings.  

We extensively evaluated MRSch under various configurations,  including cases of resource contention for either the CPU or burst buffer, by generating five synthetic workloads from the original trace (Table \ref{tab:workload}). These designed workloads represent \emph{light to heavy contentions} for the burst buffer. The assigned burst buffer request is randomly selected from the original requests within a certain range. Those greater than 5 TB are randomly assigned to S1 or S2, while S3 and S4 select from requests greater than 20 TB. Compared to S1 and S2, S3 and S4 have larger burst buffer requests. S1 and S2 have similar distributions, but more jobs in S2 include burst buffer requests. A similar pattern is observed in S3 and S4. The S5 workload is generated by reducing the requested number of nodes from S4 by half to represent workloads with less CPU resource contention. 

We split the five-month log into three parts: the first three and a half months of the workload for \emph{agent training}, a subsequent two weeks of the workload for \emph{model validation}, and the remaining data for \emph{inference/testing}.

\subsection{Evaluation Metrics} \label{sec:metrics}

The quality of the scheduling method must be evaluated
by multiple metrics, including both system-level and user-level metrics. Four well-established metrics are used to evaluate MRSch, where the first two are system-level metrics and the last two are user-level metrics.

\begin{enumerate}
\item \emph{Node utilization}: the ratio of the used node-hours during useful job execution to the elapsed node-hours.

 \item \emph{Burst buffer utilization}: the ratio of the used burst buffer hours to the elapsed burst buffer hours.

 \item \emph{Average job wait time}: the average interval between job submission to job start time. 
 
 \item \emph{Average job slowdown}: the average ratio of job response time (job runtime plus wait time) to the actual runtime, representing the responsiveness of a system.
\end{enumerate}



\begin{figure*} 
    \centering
  \subfloat[Node utilization]{%
       \includegraphics[width=0.25\linewidth]{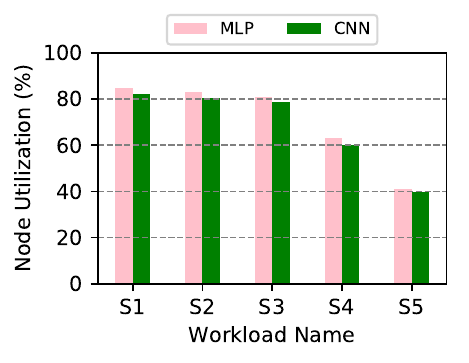}}
    \label{2a}\hfill
  \subfloat[Burst buffer utilization]{%
        \includegraphics[width=0.25\linewidth]{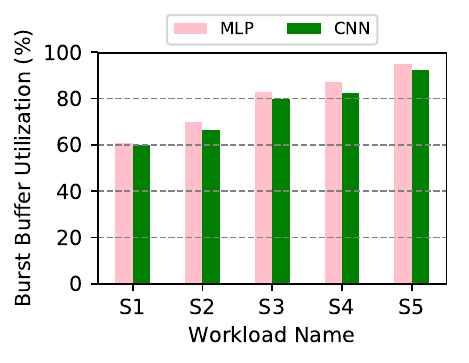}}
        \label{3a}\hfill
  \subfloat[Average job wait time]{%
       \includegraphics[width=0.25\linewidth]{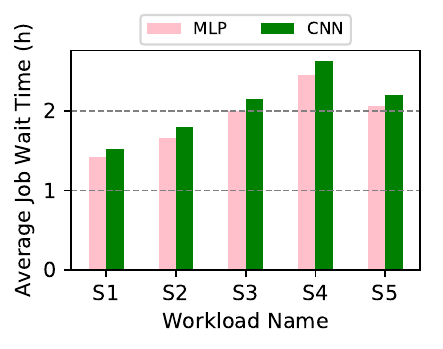}}
    \label{2a}\hfill
  \subfloat[Average job slowdown]{%
        \includegraphics[width=0.25\linewidth]{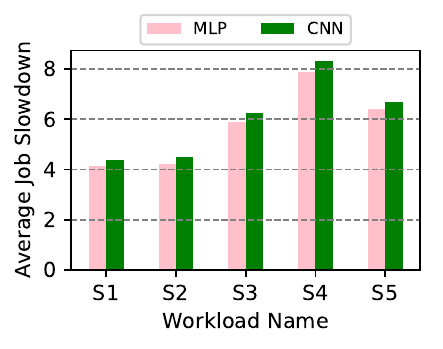}}
  \caption{Comparison of MRSch scheduling performance by using different state modules (MLP vs CNN) that indicates the use of MLP is more beneficial for multi-resource scheduling. }
  \label{fig:neuralnetwork} 
 \vspace{-0.3cm}
\end{figure*}
\begin{figure}
\centering
\includegraphics[height=1.5in,width=2.8in]{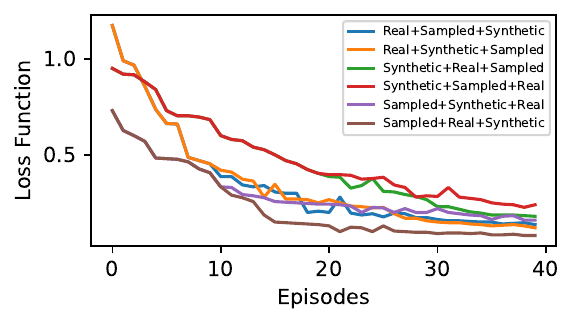}
\vspace{-10pt}
\caption{Comparison of the quality and convergence of MRSch by training with different jobset orderings. The loss function is expressed by the mean squared error.}
\vspace{-10pt}
\label{fig:training}
\end{figure}

\subsection{Network Architecture} \label{sec:parameter setting}

The input of the state neural network is a vector of size  [4$W$+2$N_{1}$+2$N_{2}$, 1], where $W$ is the window size (10 in our experiment), $N_{1}$ is the number of compute nodes, and $N_{2}$ is the number of burst buffer units in the system. For the Theta machine, the input size of the state neural network is [11410, 1]. We use two fully-connected layers with 4,000 and 1,000 neurons, respectively, with an output layer of 512 nodes. A three-layer fully-connected network with 128 neurons parses the measurement and goal modules. The action space includes the waiting jobs in the window. MRSch selects the jobs from this window for job allocation to optimize the goal. The MRSch agent follows an $\epsilon$-greedy policy to select jobs in the training time by acting greedily according to the current goal with probability $(1$ $-$ $\epsilon)$ and selects a random action with probability $\epsilon$. We set $\epsilon$ = 1.0 at the beginning of the training, which then decays at a rate of $\alpha$ = 0.995. During testing, the agent observes the environment and dynamically changes the weights in the goal vector according to the scarcity of resources calculated with Equation (\ref{normalizer}).

\subsection{Comparison Methods} \label{comparison method}
We compare MRSch with three scheduling methods:
\begin{itemize}
    \item \textbf{Heuristic} is an  extension of FCFS, belonging to the list scheduling family \cite{sun2018scheduling}, for multi-resource scheduling where jobs are scheduled according to the arrival order into the waiting queue.
  \item \textbf{Optimization} denotes the method that formulates the multi-resource scheduling problem into a multi-objective optimization problem and solves the problem using a genetic algorithm \cite{fan2019scheduling}. For a fair comparison, we apply the same window size as in MRSch.
  \item \textbf{Scalar RL} represents a group of reinforcement learning methods that formulates multi-resource requirements into a scalar reward with a fixed weight. In our experiments, we use a policy gradient method \cite{silver2014deterministic} and the scalar reward is set to (0.5$\times$\emph{CPU\_util} $+$ 0.5$\times$\emph{burst buffer\_util}).
\end{itemize}

 In addition, EASY backfilling is adopted in each of these methods  to mitigate resource fragmentation \cite{mu2001utilization}. The comparison study is performed with the trace-based, event-driven scheduling simulator CQSim \cite{CQSim}.

\begin{figure*} 
    \centering
  \subfloat[Node utilization]{%
       \includegraphics[width=0.41\linewidth]{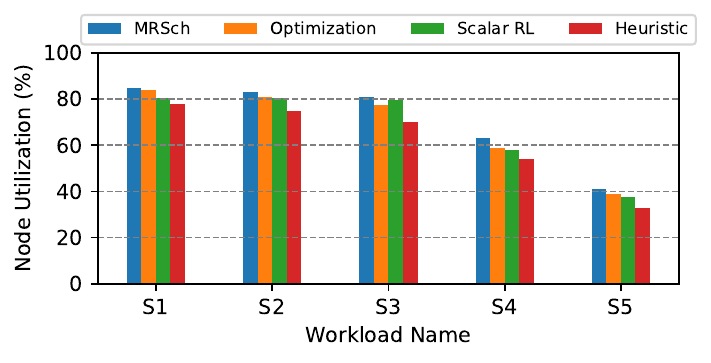}}
    \label{3a}\hspace*{0.06\textwidth}
  \subfloat[Burst buffer utilization]{%
        \includegraphics[width=0.41\linewidth]{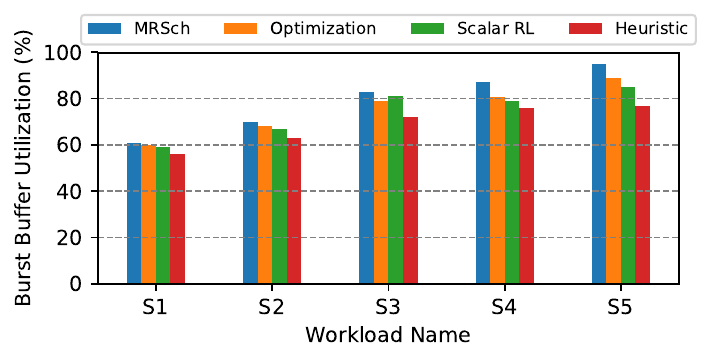}}
    \label{3b}

  \caption{Scheduling performance in terms of system-level metrics.}
  \vspace{-10pt}
  \label{fig:sytem_perf} 
\end{figure*}

\begin{figure*} 
    \centering
  \subfloat[Average job wait time]{%
       \includegraphics[width=0.41\linewidth]{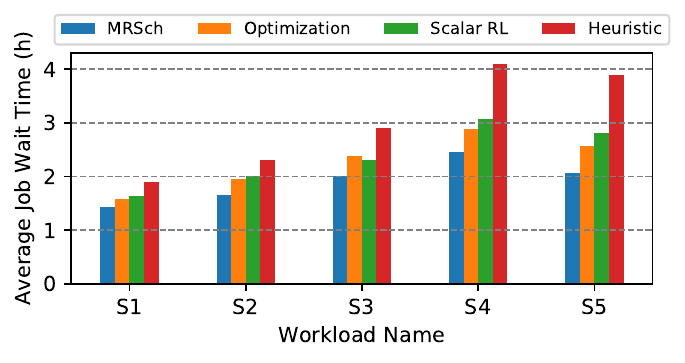}}
    \label{3a}\hspace*{0.06\textwidth}
  \subfloat[Average job slowdown]{%
        \includegraphics[width=0.41\linewidth]{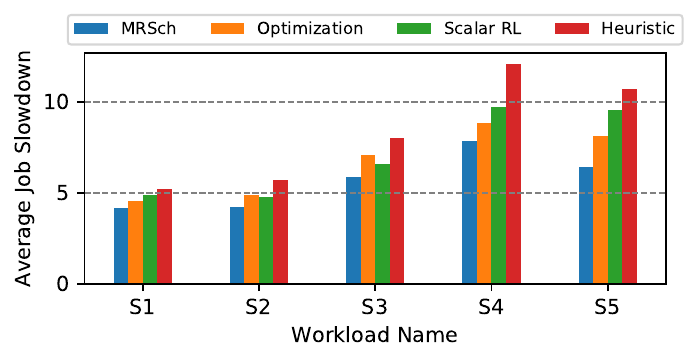}}
    \label{3b}

  \caption{Scheduling performance in terms of user-level metrics.}
  \vspace{-10pt}
  \label{fig:user_perf} 
\end{figure*}

\section{Results} \label{sec:result}

We examine the scheduling performance of MRSch under different state module representations and various training strategies in \S \ref{network performance} and \S \ref{training performance}, and compare MRSch with existing multi-resource scheduling methods in \S \ref{system performance}. We also assess if MRSch can adapt to workload changes in \S \ref{adaption}. A case study with more schedulable resources is presented in \S \ref{sec:3 resources}. Finally, we list runtime overhead in \S \ref{sec:overhead}.

\subsection{State Module: MLP vs CNN} \label{network performance}

For the state module described in \S \ref{sec:state representation}, we use MLP instead of the CNN adopted in the original DFP. This set of experiments compares 
the scheduling performance under different state modules (MLP vs CNN), with results presented in Figure \ref{fig:neuralnetwork}. The use of the MLP network achieves a better scheduling performance by up to 7\% across the system-level and user-level metrics. CNN is good at spatially extracting local correlations present in the input data, e.g., local filters over the input \cite{alzubaidi2021review}. However, the features of the MRSch state module input (e.g., 
the waiting job and running job information) are not spatially related. For processing input with independent features, MLP often provides a better solution \cite{goodfellow2016deep}, as we observe in this experimental scenario.

\subsection{Training Strategy} \label{training performance}

We separate the training data into ten job sets and collect another ten job sets by randomly sampling jobs from the original training trace. The arrival times for these jobs are modeled as a Poisson distribution following the average inter-arrival time of the original trace. Also, we generate 20 synthetic job sets that mimic Theta workload patterns in terms of hourly and daily job arrivals, distributions of resource requests, and job runtimes. In total, we train our model with 40 job sets containing 200,000 jobs.

Figure \ref{fig:training} compares the convergence rates for the loss function by varying the order of the job sets during the training of MRSch.  
This set of experiments demonstrates that training with a sequence of sampled, real, and synthetic job sets (the brown curve in the figure) achieves the fastest convergence speed and the smallest mean squared error compared to the other trace orderings. 
This result  confirms our initial intuition that it is advantageous for MRSch to first learn from simple, averaged cases (i.e., the sampled job sets) and then subsequently advance through more complex special cases
(i.e., the real and synthetic job sets) to generate a converged and high-quality model.

\subsection{Scheduling Performance} \label{system performance}

 Figure \ref{fig:sytem_perf} compares different scheduling methods in terms of the system-level metrics.  MRSch yields the highest node and burst buffer utilization across the various  workloads, whereas FCFS leads to the worst system performance. 
 Scalar RL achieves better performance than the optimization method on S3, which we attribute to two reasons. First, compared to the other workloads, the CPU and burst buffer demands in S3 are relatively balanced. Second, the RL method offers better long-term effects due to its learning capability \cite{fan2021deep}.

Among the five workloads, MRSch achieves a larger increase in resource utilization on  S4 and S5. The contention fierceness for the burst buffer increases from S1 to S5, which indicates that MRSch attains higher performance gains when the resource demands are unbalanced and fierce.

Figure \ref{fig:user_perf}  compares different scheduling methods in terms of the user-level metrics. For all cases, MRSch achieves the best performance. We notice that average job wait time and slow down
increase dramatically as the burst buffer requests increase.  The most noticeable average wait time and slowdown reductions obtained by MRSch occur in the heavily unbalanced S4 and S5 workloads. In contrast, the scalar RL method does not perform well in these workloads, suggesting the importance of dynamic resource prioritization in response to the scarcity of the resources. These results also indicate that MRSch achieves better scheduling performance, especially in the cases of high demand from user jobs for a resource. Overall, MRSch delivers the best performance among all compared methods, highlighted by shortening the average job wait time by up to 48\% and decreasing job slowdown by up to 41\%.

\begin{figure*}[!htbp]
\centering
\subfloat[S1 workload.]{\includegraphics[width=0.28\textwidth]{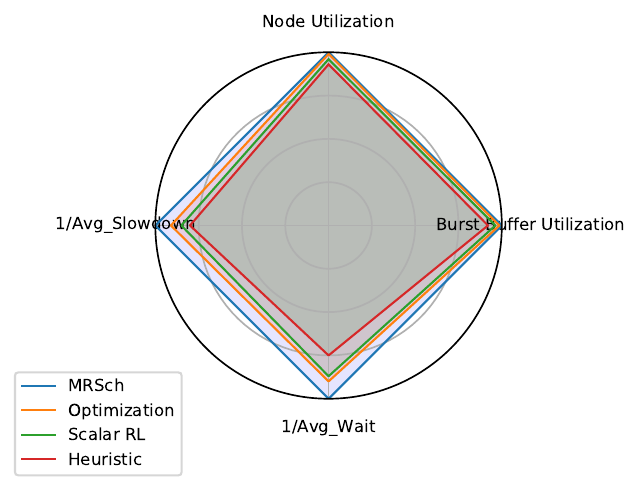}}\hspace*{0.02\textwidth}%
\subfloat[S2 workload.]{\includegraphics[width=0.28\textwidth]{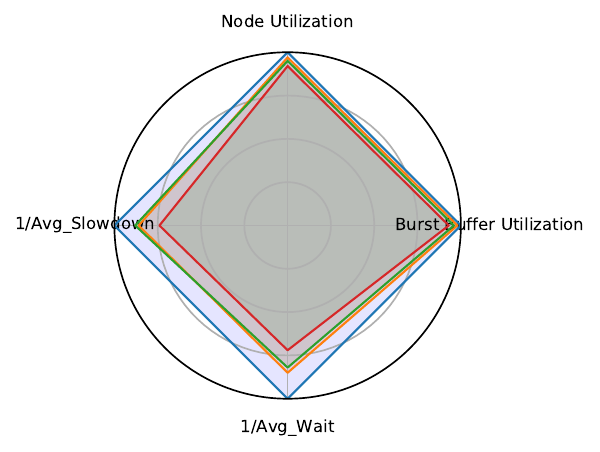}}\hspace*{0.02\textwidth}%
\subfloat[S3 workload.]{\includegraphics[width=0.28\textwidth]{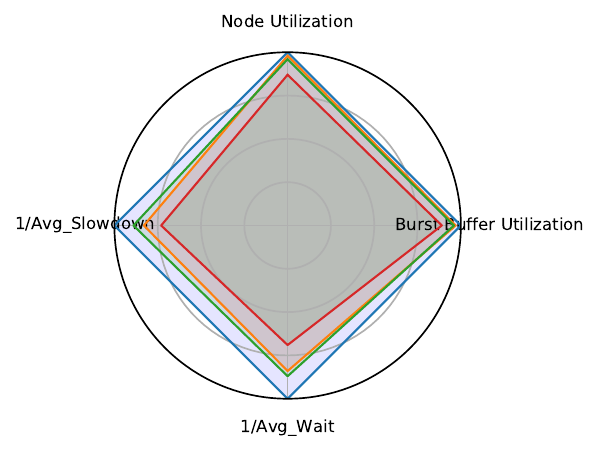}}

\subfloat[S4 workload.]{\includegraphics[width=0.28\textwidth]{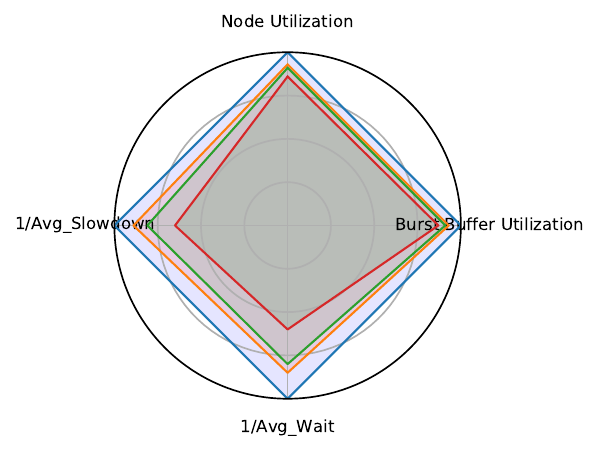}}%
\hspace*{0.02\textwidth}%
\subfloat[S5 workload.]{\includegraphics[width=0.28\textwidth]{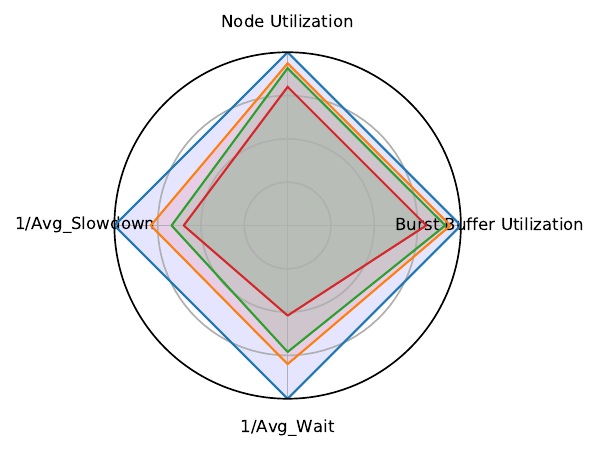}}
  \caption{Comparison of MRSch with existing methods using Kiviat charts. 
The larger the area outlined by a curve, the better the overall performance of the corresponding method.}
  \label{fig:radar_all} 
\end{figure*}

\begin{figure}
\centering
\includegraphics[height=1.1in,width=3in]{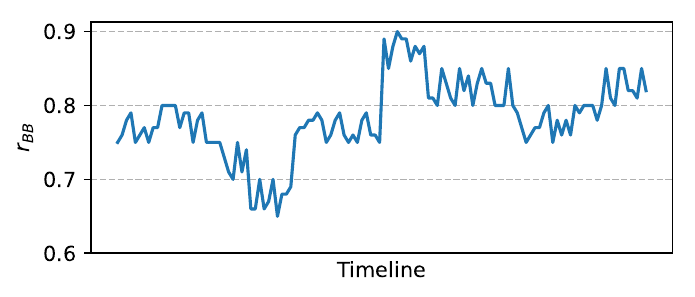}
\vspace{-10pt}
\caption{The fluctuation of $r_{BB}$ under the S5 workload within a randomly selected 12 hours, where $r_{BB}$ is calculated by Equation (\ref{normalizer}). This value reflects the relative importance of  burst buffer compared to CPU during multi-resource scheduling.}
\label{fig:s5}
\end{figure}

\begin{figure}
\centering
\includegraphics[height=1.1in,width=2in]{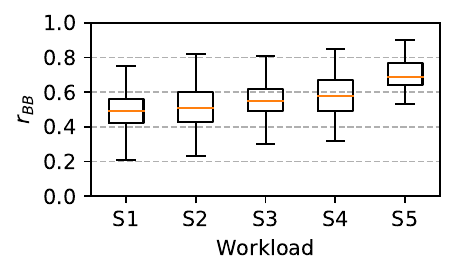}
\vspace{-10pt}
\caption{Box plot of $r_{BB}$ (i.e., the goal vector value corresponding to the burst buffer utilization) for the S1--S5 workloads.}
\vspace{-10pt}
\label{fig:burst buffer}
\end{figure}

\begin{figure*}[!htbp]
\centering
\subfloat[S6 workload.]{\includegraphics[width=0.28\textwidth]{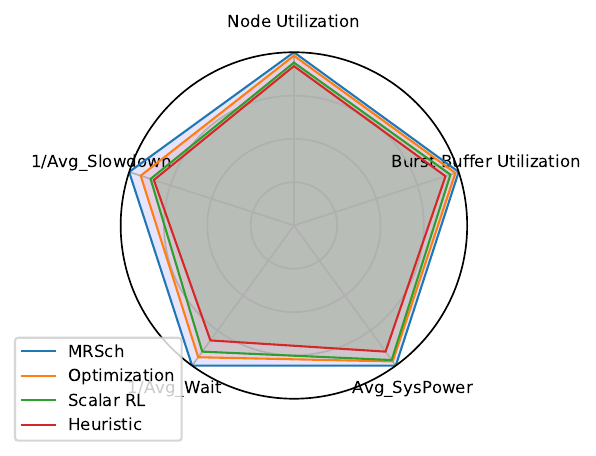}}\hspace*{0.02\textwidth}%
\subfloat[S7 workload.]{\includegraphics[width=0.28\textwidth]{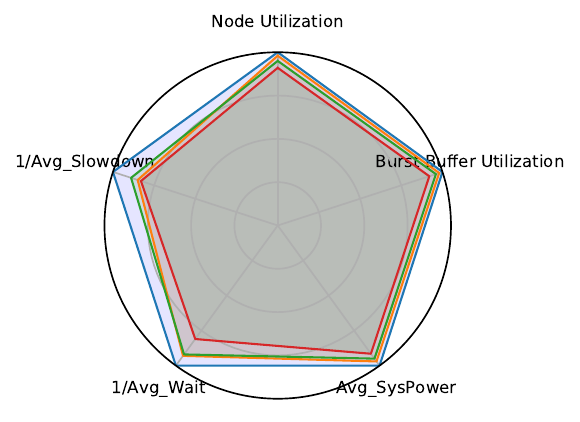}}\hspace*{0.02\textwidth}%
\subfloat[S8 workload.]{\includegraphics[width=0.28\textwidth]{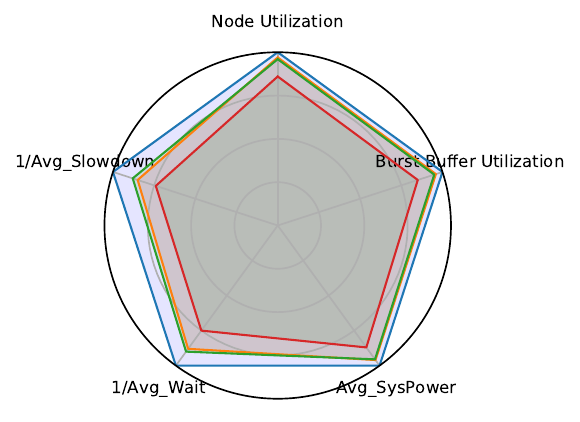}}

\subfloat[S9 workload.]{\includegraphics[width=0.28\textwidth]{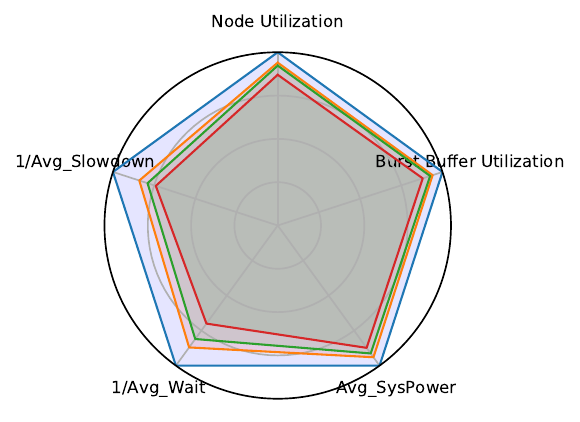}}%
\hspace*{0.02\textwidth}%
\subfloat[S10 workload.]{\includegraphics[width=0.28\textwidth]{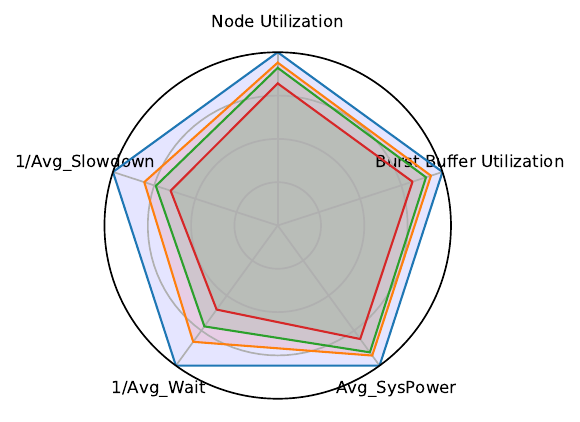}}
  \caption{Scheduling performance under more schedulable resources.
The larger the area outlined by the curve, the better the overall performance of the corresponding method.}
  \label{fig:radar_power} 
\vspace{-0.5cm}
\end{figure*}

Figure \ref{fig:radar_all} presents Kiviat charts of the overall scheduling performance for each workload obtained by different scheduling methods. We plot the reciprocal of the average job wait time and the reciprocal of the job slowdown in these charts. All metrics are normalized within the range of $[0,1]$, where $1$ corresponds to a method that achieves the best performance among all others. In other words, a larger area outlined in the figure indicates a better overall scheduling performance for that method. MRSch consistently yields the best results, whereas  FCFS delivers the worst performance across all the workloads.  

MRSch demonstrates its best improvements over the other methods in S5. We attribute this outcome to the heavily unbalanced contention for each resource compared to the S1--S4 workloads. Specifically, the burst buffer resource contention in S5 is more fierce than the CPU resource contention. MRSch delivering its strongest performance in this scenario suggests its robust capability to automatically change objectives within an unbalanced resource contention environment.

\vspace{-0.5em}

 \subsection{Adaption to Workload Change} \label{adaption}
To validate our observations, we examine $r_{BB}$, the goal vector value for the burst buffer as calculated by Equation (\ref{normalizer}). This value reflects the relative importance of the burst buffer to CPU during multi-resource scheduling.
Figure \ref{fig:s5} plots the dynamic changes of $r_{BB}$ in the range of  0.6 to 0.9 when using MRSch in  S5 during a randomly selected 12 hours. 

Figure \ref{fig:burst buffer} presents box plots of $r_{BB}$ in the S1--S5 workloads, suggesting that (1) $r_{BB}$ dynamically changes compared to the fixed value of $0.5$ in the scalar RL method, and (2) the minimum value, first quartile, mean value, third quartile, and maximum value are the largest for S5. These results validate that the MRSch agent automatically assigns more preference to the relatively scarce resources when it detects an unbalanced contention for each resource. However, in such a situation, the  scalar RL method treats the CPU and burst buffer equally, which leads to poor scheduling performance. 

\subsection{Case Study: More Resources} \label{sec:3 resources}

MRSch is generally applicable for multiple schedulable resources.  For instance, another schedulable resource could be power because the power consumption of supercomputers increases significantly. Aurora, the planned exascale supercomputer, anticipates a power budget of 60 MW \cite{Aurora}. Therefore, approaches for improving energy efficiency are attracting more attention in the HPC field, including several studies that explored power-aware scheduling \cite{wallace2016data,borghesi2018scheduling}. As a case study, we incorporate power as a third resource in addition to the CPU and burst buffer resources to illustrate how MRSch can be easily extended to incorporate more schedulable resources.

Consider a system with three schedulable resources of CPU, burst buffer, and power. A fixed power budget exists for the entire system, making power another resource for which jobs must contend.  Each submitted job includes four pieces of information: walltime, requested number of nodes, requested volume of burst buffer, and a power profile (i.e., peak power consumption). This case study considers three objectives of maximizing the CPU/node utilization, maximizing the burst buffer utilization, and maximizing the total power consumption of running jobs within a power budget. 

We generate five new workloads S6--S10 by creating power profiles for the jobs from S1--S5.  For each job, its power consumption per node is randomly assigned between (100--215 W). The Theta computing nodes (Intel KNL 7230) have a 215 W thermal design power (TDP) \cite{TDP}, and 100 W is selected as the lower bound based on previous work \cite{sharmadynamic}. The power consumption of an idle node is set to 60 W \cite{marincic2017polimer}, and the power budget for the entire system is restricted to 500 kW to ensure a contentious environment.

Figure \ref{fig:radar_power} presents a holistic view of the scheduling performance. We observe that MRSch achieves the best overall performance for all workloads, while the FCFS heuristic results in the worst overall performance on all workloads. Compared to the other methods, MRSch improves resource utilization by up to 18\%, reduces the average wait time by up to 39\%, and reduces the average slowdown by up to 34\%, which demonstrate that MRSch can be generally applied to scheduling multiple resources. In summary, this case study demonstrates the effectiveness of MRSch for multiple schedulable resources.  

\vspace{-0.25mm}

\subsection{Runtime Overhead} \label{sec:overhead}

In our experiments, MRSch required less than two seconds to make scheduling decisions during the two-resource scheduling experiments and less than three seconds during the three-resource scheduling during testing. All experiments were performed on a personal computer configured with an Intel 2 GHz quad-core CPU and 16 GB memory. Current HPC systems typically require the scheduler to respond within 15--30 seconds \cite{yang2013integrating}. Therefore, the MRSch agent imposes negligible overhead and is a feasible solution for online deployment in production systems.

\section{Conclusion} \label{conclusion}
Motivated by the increasing need for multi-resource scheduling in HPC, we present MRSch, an intelligent multi-resource scheduling agent that leverages an advanced multi-objective reinforcement learning algorithm called DFP. While DFP features an inherent advantage for pursuing dynamically changing objectives, it was initially designed for gaming and never previously applied to HPC scheduling. In this work, we describe our problem formulation and several key techniques are developed into MRSch for incorporating HPC-specific scheduling requirements. 
These techniques enable MRSch to automatically observe the HPC scheduling environment and adapt its policy to continuous workload and resource changes.
Our experimental results show that MRSch outperforms existing scheduling approaches---heuristic, optimization, and scalar-based reinforcement learning methods---by up to 48\%  in terms of user-level and system-level metrics. 

While MRSch demonstrates promising performance compared with  conventional heuristic and  optimization methods, a significant gap remains in deploying RL-based scheduling in production systems. One key hurdle is the lack of model interpretability. Because the scheduling agent is constructed on deep neural networks with millions or more parameters, it appears as a  black box model to system managers, so is incomprehensible to debug, deploy, and adjust in practice \cite{Meng2020}. Our future work includes investigating how to provide practical RL-driven scheduling systems with interpretable models. 


\section*{Acknowledgment}
This work is supported in part by US National Science Foundation grants CNS-1717763, CCF-2109316, CCF-
2119294, and U.S. Department of Energy, Office of Science, under contract DE-AC02-06CH11357.

\bibliographystyle{IEEEtran}
\bibliography{bib/cluster.bib}
\end{document}